\documentclass[twocolumn,preprintnumbers,amsmath,amssymb]{revtex4}

\usepackage{amsmath}

\begin{document}

\preprint{APS/123-QED}

\title{Talk about pivots}

\author{Damien Martin}
\author{Andreas Albrecht}%
 \email{djmartin@ucdavis.edu}
 \email{albrecht@physics.ucdavis.edu}
\affiliation{Department of Physics, University of California, Davis, CA 95616, USA}

\date{\today}

\begin{abstract} 

We consider two different methods of
parameterizing the dark energy equation of state in order to assess
possible ``figures of merit'' for evaluating dark energy experiments.
The two models are $w(a) = w_0 + (1-a)w_a$ and
$w(a) = w_p + (a_p - a)w_a$.  This brief note shows that
the size of the error contours cannot change under such a
reparameterization. This makes the figures of merit associated with
these parameterizations mathematically identical. This also means that
any ``bias'' exhibited by one model is equally present in the other.
\end{abstract}


\maketitle

\section{Introduction}
Dark energy poses challenges to physics because we know so little
about it. While there are many competing models, none of them are
overwhelmingly convincing. This is of course why the field is so
exciting.  We expect to learn a great deal by further study of the
dark energy. This note addresses a specific technical issue related to
modeling dark energy and forecasting the impact of future experiments. 

As discussed in \cite{Linder:2006xb} it is standard practice to model
the dark energy as a perfect fluid.  Its dynamical properties can then
be expressed in terms of the
the dark energy equation of state parameter $w$ as a function
of cosmic scalefactor $a$. The general case of this
parameterization implies an infinite number of degrees of freedom (in 
order to describe the continuous function $w(a)$), but in most
implementations a finite parameter ansatz is used. Any simple ansatz
will exclude possible forms for $w(a)$ and thus could distort the
subsequent analysis, so the relative advantages and risks associated
with particular choices of ansatz are a topic of ongoing discussion
and debate.

Currently, the ansatz that is probably most commonly used
is the linear expression
\begin{equation}
w(a) = w_0 + (1-a)w_a \label{eq:normal}
\end{equation}
which we will call ``normal form''. This is a special ($a_p = 1$) case of the linear expression
\begin{equation}
w(a) = w_p + (a_p-a)w_a\label{eq:pivot}
\end{equation}
where $a_p$ is the ``pivot scalefactor''. Equation \eqref{eq:pivot} is
usually used as in \cite{Hu:2003pt}, where $a_p$ is chosen to be the
value of the 
scalefactor where $w(a)$ is most tightly constrained by a
particular data set \cite{Huterer:2000mj}.  By construction, $w(a_p)
\equiv w_p$ in this 
ansatz, which we will call ``pivot form''.  

Attention has been recently drawn to the pivot ansatz by the 
Dark Energy Task Force (DETF) \cite{Albrecht:2006}. 
The DETF defines a figure of merit for a given data set based on the
area inside a constant probability contour in the two dimensional
$w_p$--$w_a$ plane. 

In \cite{Linder:2006xb} Linder looked
at the effect of moving from a description based on normal form 
to a description based on pivot form (\S III
of \cite{Linder:2006xb}). 
Linder's paper suggests that the normal form
suffers less from bias than the pivot form. He
considers a non-linear generalization of \eqref{eq:pivot} where $a$ has an
exponent $b$ differing from unity, and demonstrates a bias using this
generalized pivot form in his figures 2 and 3.

This brief note focuses on the {\em linear} ($b=1$) form of the pivot ansatz.
This is the only form that is relevant to the DETF work.  We show
why normal form and pivot form are mathematically identical and
interchangeable when it comes to defining ``area'' figures of merit
like the one used by the DETF (this fact has already been stated in
the Fisher matrix 
approximation in \cite{Albrecht:2006}.)  Linder's discussion emphasizes the 
bias he exhibits using his non-linear ansatz and singles out the
behavior of $w_p$.  Linder's discussion favors
normal form over pivot form, stating that normal form is ``more robust''.

While it may seem that our conclusion is in conflict with
\cite{Linder:2006xb}, we do not believe there are any concrete technical
points of disagreement.  For example, Linder's figures 2 and 3 shows
that the bias disappears for the linear ($b=1$) case.  Also, we agree
with Linder that the choice of $a_p$ is data-dependent.  So the 
differences between this comment and \cite{Linder:2006xb} are
apparently  only
ones of emphasis.  Our message is that 
\cite{Linder:2006xb} does not demonstrate any weakness
of the DETF figure of merit relative to the equivalent (in fact equal)
one based on normal form. Our motivation is to make this message
transparent even in the non-Gaussian case.

\section{A mechanical analogue}
As the scalefactor plays the role of the evolution parameter, it is
useful to think of it as time. Identifying the equation of state
$w(a)$ as $x(t)$ we transform \eqref{eq:normal} and \eqref{eq:pivot}
into the equation of a particle traveling at a constant velocity: 
\begin{equation}
x(t) = x_0 - t x_a
\end{equation}
It is well known that this mechanical system comes from an almost
trivial Hamiltonian  
\begin{equation}
H = \frac{p^2}{2}
\end{equation}
where the ``mass'' has been scaled to unity. As this is a
time-independent evolution of the system we can apply Liouville's
theorem: the evolution of a region of phase space is volume
preserving. That is, time evolution preserves the volume in the
$x_0$--$x_a$ plane. 

\section{Comments}
The difference between Eqn. \eqref{eq:normal} and Eqn. \eqref{eq:pivot} is that
they are looking at the same system at different ``times'' or scalefactors. As these equations have a Hamiltonian evolution (as is easy
seen by constructing a mechanical analogy) the phase space volume
cannot change. As a result, the area inside the error contours cannot
depend on whether you are using parameterization \eqref{eq:normal} or
\eqref{eq:pivot}, although the shape and orientation of the contours
in $w_0$--$w_a$ space can certainly change. 
This result is interesting because we have not
made any assumption about the underlying distribution in the
$w_0$--$w_a$ plane. (The case of a Gaussian distribution is considered
explicitly in the Appendix.)  Both
pivot and normal form consider the exact same linear family of functions
$w(a)$, and just label them by different linear combinations of their
parameters. This particular relabeling does not change area in phase
space, and does not change the likelihood of a specific function $w(a)$
given a specific data set.

More generally, it is interesting to ask what happens when the class
of models is generalized so that the coefficient of $a$ is no longer
unity. In this case, it is \emph{no longer true} that the area in the
$w_0$--$w_a$ plane is preserved. The main reason why the above
explanation fails is that the ``canonical momentum'' changes in time,
and while the ``$q$--$p$'' area \emph{is} preserved in time, the
$w_0$--$w_a$ area is not.   However, this generalization is not relevant
to normal and pivot forms, and thus does not impact our conclusions.

So we emphasize again our main point: The area figures of merit
based on pivot and normal form are mathematically identical.

\section{Acknowledgements}
We thank Lloyd Knox and Eric Linder for helpful comments.  This work
was supported in part by DOE grant DE-FG03-91ER40674.

\appendix

\section{The Fisher matrix case}
We have shown generally the equivalence of area figures of merit
based on the pivot and normal forms.  Here we 
show explicitly how this result comes out of the Fisher matrix analysis
of a Gaussian distribution probability distribution $P(\mathbf x)$ as a
function of parameter vector ${\mathbf x}$. 

Here we have 
\begin{equation}
P(\mathbf x) = \frac{1}{\sqrt{2\pi\det \mathbf{F}}} \exp(-\mathbf{x}^T \mathbf{F} \mathbf{x}).
\end{equation}
where $\mathbf{F}$ is the covariance matrix, meaning that the
eigenvectors give the directions of the principal axes of the error
ellipse and the eigenvalues give the inverse variances squared ($1/\sigma_1^2$
and $1/\sigma_2^2$) in these
directions, as is easily seen by diagonalizing $\mathbf{F}$. The
figure of merit $\sigma_1 \times \sigma_2$ is, up to a constant factor, 
the area of the 1-$\sigma$ ellipse. 

When we change basis to $\mathbf{x}^\prime = \mathbf{Tx}$ then the
quadratic form transforms in the following way: 
\begin{equation}
(\mathbf{x}^{\prime})^T(\mathbf{T}^T \mathbf{FT})\mathbf{x}^\prime \equiv 
(\mathbf{x}^{\prime})^T\mathbf{F}^\prime \mathbf{x}^\prime
\end{equation}
where $\mathbf{F}^\prime$ is the transformed covariance matrix. In our case, we have
\begin{equation}
{\mathbf x^\prime} = 
\left(\begin{array}{c}
w_p\\
w_a
\end{array}
\right)
 = 
\left(\begin{array}{cc}
 1 & 1-a_p\\
 0 & 1
\end{array}
\right)
\left(\begin{array}{c}
w_0\\
w_a
\end{array}
\right),\quad
\mathbf{x} = 
\left(\begin{array}{c}
w_0\\
w_a
\end{array}
\right)
\end{equation}  
as can easily be verified. The transformation matrix has unit determinant, and so
\begin{equation}
\frac{1}{(\sigma_1^\prime)^2} \times \frac{1}{(\sigma_2^\prime)^2} = \det
\mathbf{T}^T\mathbf{FT} = \det \mathbf{F} = \frac{1}{\sigma_1^2}
\times \frac{1}{\sigma_2^2} 
\end{equation}
which shows that the figure of merit could not have changed under this transformation. 

It should be emphasized that this is not a separate result, but rather
a special case of the result in the body of this note presented in a
language that may be more familiar to some readers.

\end{document}